\documentclass[aps,pra,10pt,twocolumn,superscriptaddress]{revtex4}

\usepackage{amsmath}
\usepackage{amssymb}
\usepackage{bbm}
\usepackage{graphicx}
\usepackage{framed}
\usepackage{color}
\usepackage[colorlinks = true, urlcolor = blue]{hyperref}
\usepackage{epstopdf}
\usepackage{dsfont}
\usepackage{amssymb}
\usepackage{amsmath}
\usepackage{amsthm}
\usepackage{wrapfig}
\usepackage{relsize}
\usepackage{bm}
\usepackage[latin1]{inputenc}
\usepackage{enumitem}
\usepackage{natbib}
\usepackage{graphicx}
\usepackage{physics}
\usepackage{hyperref}
\usepackage{xcolor}

\begin{document}

	\title{Quantum  capacities bounds
	in spin-network communication channels} 

	\author{Stefano Chessa}
	\email{stefano.chessa@sns.it}
	\author{Marco Fanizza}
	\author{Vittorio Giovannetti}
	\affiliation{NEST, Scuola Normale Superiore and Istituto Nanoscienze-CNR, I-56126 Pisa, Italy}
	\date{\today}

\begin{abstract}
Using the Lieb-Robinson inequality and the continuity property of the quantum capacities in terms of the diamond norm, we derive an upper bound on the values that these capacities can attain in spin-network communication models of arbitrary topology. Differently from previous results
we make no assumptions about the encoding mechanisms that the sender of the messages adopts in loading information on the
network. 
\end{abstract}

\maketitle

\section{Introduction} \label{sec.Intro}

In the flying qubit model of quantum communication messages are conveyed from the sender (Alice) to the intended receiver (Bob) after being encoded into 
some degree of freedom which actually ``moves" from the location of the first party to the location of the second party~\cite{HOLEVOBOOK,WILDEBOOK,WatrousBOOK}. 
This scenario is the most widely studied in the literature as it finds application in many realistic scenarios which, for instance, employ electro-magnetic pulses 
as  quantum carriers.   
An intriguing alternative is provided by the spin-network communication (SNC)  
model where instead Alice and Bob are assumed to have access to different portions of an extended  many-body quantum medium
formed by interacting particles which occupy fixed locations but which are mutually coupled via an assigned, fixed Hamiltonian that, as in a solid,  allows the spread of local perturbations along the medium,  see e.g. Ref.~\cite{BOSE1} and references therein.
While being intrinsically limited to short distance applications, SNC schemes have been suggested as an effective way to avoid interfacing issues in the 
 engineering  of connections between clusters of otherwise independent quantum processors~\cite{BOSE,SPIN1,SPIN2,
SPIN3,SPIN4,SPIN5,SPIN6,DANIEL}. 
The study of these models is also motivated by the need of better understanding how 
the many-body system reacts to the spreading of local perturbations. 
The main result in this context is the well known bound by  
{Lieb and Robinson} (LR)~\cite{LR,REVIEW} on the maximum group velocity for two-points correlation functions
of the network, see also~\cite{Clust Theo,LSM Theo,exponential1,ExistDynam}.
For sufficiently regular models, it basically 
identifies the presence of an effective light cone with exponentially decaying tails
implying that information that leaks out to space-like separated regions is negligible, so that for large enough distances  non-signaling is preserved.
Several applications of the LR inequality in a quantum information theoretical treatment of SNC models have been presented in the literature. For instance in 
Ref.~\cite{JENS} the LR bound was used to set a limit on the entanglement that can develop across the boundary of a distinguished region for short times. In Ref.~\cite{OSB} instead the bound was used to show that 
dynamics of 1D quantum spin systems can be approximated efficiently. 
In  Ref.~\cite{C1} finally,  
 making use of the Fannes inequality~\cite{FANNES}, Bravyi {\it et al.}  succeeded in linking 
 the LR inequality  to the 
 Holevo information capacity $C_1$~\cite{CAP1,CAP2} attainable for a special example of  SNC model where
 Alice tries to communicate classical messages to Bob by ``overwriting" them into the initial state of the spin-network she controls.  A  generalization of this result was presented in  Ref.~\cite{EPSTEIN} where the LR
 bound was employed to set the limits within which 
high-fidelity quantum state transfer and entanglement generation can be performed in general spin-network systems. 
\begin{figure}[t!]
  \includegraphics[width=\linewidth]{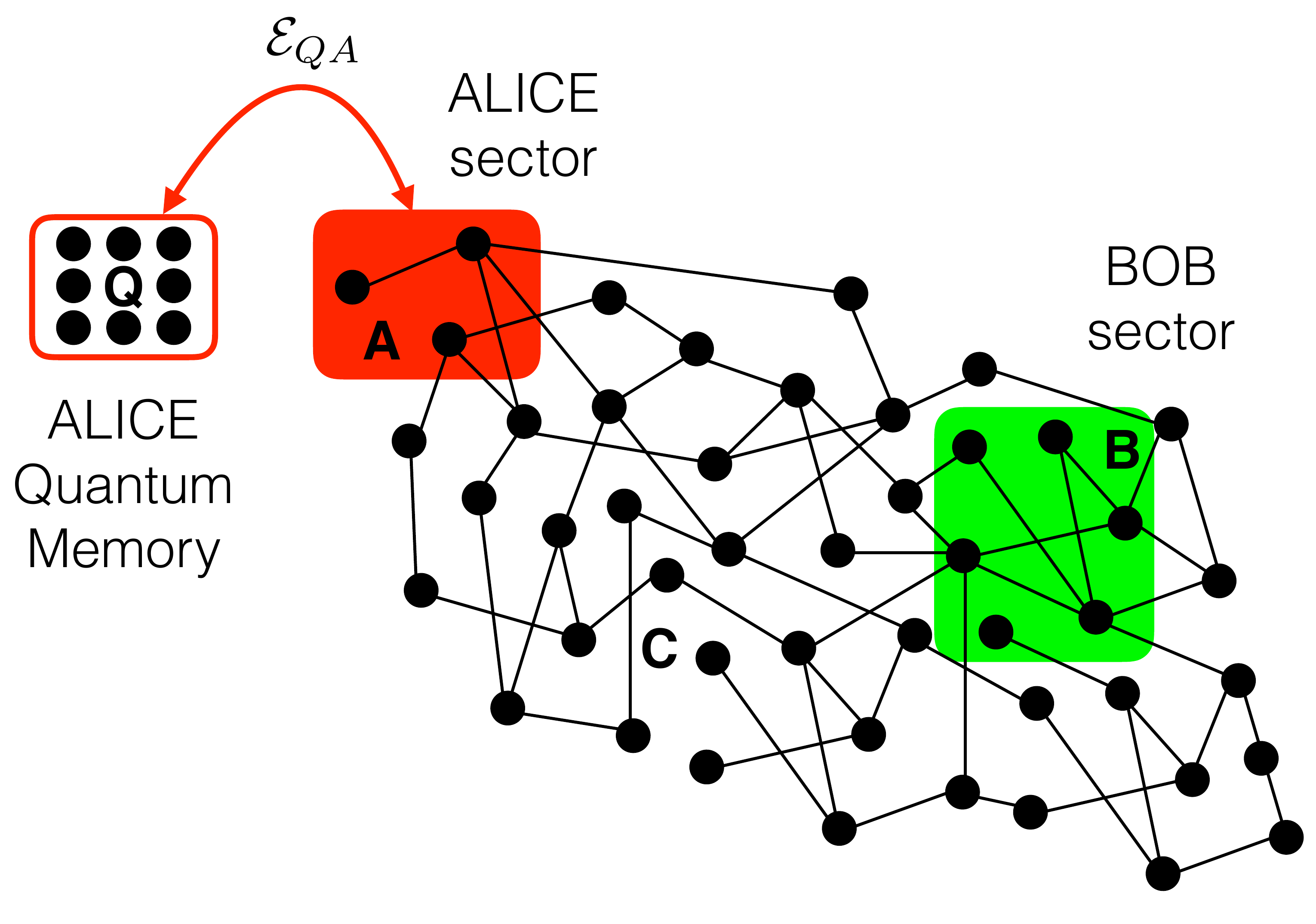}
  \caption{(Color online) Schematic representation of a spin-network model for quantum communication.
  The network ${\cal N}$ is  divided into three components: the sector $A$ (controlled by the sender of the message Alice), the sector $B$ (controlled by the receiver Bob), and the sector $C$ on which neither Alice nor Bob can operate. 
  The element $Q$ represents an external ancillary memory element Alice uses to
  store the information she wants to transmit.
  At time $t=0$ Alice couples $A$ 
 with $Q$ via an arbitrary encoding mapping ${\cal E}_{QA}$ which fully characterizes  the adopted communication strategy; Bob, on his side, will try to pick up the message
  at some later time $t$ from $B$. }
  \label{fig:network}
\end{figure}
The aim of the present manuscript is to go beyond these findings,  by 
 generalizing the inequality derived in Ref.~\cite{C1} to the whole plethora of quantum channel capacities~\cite{REFCAP} that one can associate to the underlying SNC model and to the arbitrary encoding  strategies Alice may adopt to upload her messages into the network. 
For this purpose we shall make explicit use of the the continuity argument of Refs.~\cite{LEUNG,SHIROKOV} which allows one
to connect the capacities values  of two channels via their relative distance measured in terms of the diamond norm metric~\cite{DIAMOND,DIAMOND1}.
While our derivation in many respects mimics the one presented by Bravyi {\it et al.}, we stress that in order to account for all 
possible encoding strategies, we have explicitly to deal with the  dimension of the ancillary memory element $Q$ Alice can use in the process. 
The presence of such element, which does not enter in the definition of the spin-network (and hence in the 
associated LR inequality), introduces a divergent contribution which, if not properly tamed,  tends to spoil the connection between the LR bound and the diamond norm distance, compromising the possibility of using the results of Refs.~\cite{LEUNG,SHIROKOV}
to constrain the capacities values of the underlying
SNC model (a problem which, due to the
intrinsic sub-additivity of the Holevo information  $C_1$, needed not to be addressed in Ref.~\cite{C1}).

The manuscript is organized as follows:
we start in Sec.~\ref{SEC:CAP} by introducing the SNC scheme and reviewing some basic facts about the LR bound. 
The main results of the paper are presented in Sec.~\ref{Sec:nonsig}. Here, in Sec.~\ref{tracedist}, first we exploited the LR inequality to put an upper limit on the induced trace-norm distance~\cite{WatrousBOOK} 
between the map associated with the SNC scheme and a (zero-capacity) completely depolarizing channel~\cite{DEP,NC}.
 From this, in Sec.~\ref{secdiamond} 
  we hence derive an analogous  bound for the diamond distance~\cite{DIAMOND,DIAMOND1}
  from 
  which ultimately the bounds on the SNC communication capacities follow.  The paper ends with the conclusions in Sec.~\ref{Sec:conc}. Technical material is presented in the Appendix.

\section{The model} \label{SEC:CAP}

In the scenario we are interested in, two distant parties (Alice the sender and Bob the receiver)  try to exchange (classical or quantum) messages 
by locally manipulating portions of a many-body quantum system~${\cal N}$ that, as schematically shown in Fig.~\ref{fig:network},
acts as the mediator of the information exchange~\cite{BOSE,BOSE1,SPIN1,SPIN2,SPIN3,SPIN4,SPIN5,SPIN6,C1}.
An exhaustive characterization of ${\cal N}$ is provided by the spin network formalism~\cite{Clust Theo} 
where the (fixed) locations of the quantum subsystems  are specified by a 
 graph $\mathbb{G}:=(V,E)$ defined by a set of
 vertices $V$ and by a set $E$ of edges. The model is equipped with a metric  $d(x,y)$ defined as the shortest path (least number of edges) connecting $x,y \in  V$ ($d(x,y)$ being set equal to infinity in the absence of a connecting path), which induces a measure for 
the diameter $D(X)$ of a given subset $X\subset V$, and a  distance $d(X,Y)$ between
 the subsets  $X,Y \subset V$, 
   \begin{eqnarray}  
   D(X)&:=&\max\limits_{x,y} \min \{ d(x,y) | x,y\in X\}\;, \nonumber \\ 
   d(X,Y) &:=& \min \{ d(x,y) | x \in X ,y\in Y\}\;. \end{eqnarray} 
Indicating  with ${\cal H}_x$ the Hilbert space associated with the spin that occupies the vertex $x$ of the graph, 
 the  Hamiltonian of ${\cal N}$, which  ultimately is responsible for the information propagation in the medium,  can be expressed as 
\begin{eqnarray} \label{HAMILT} 
\hat{H} : = \sum_{X\subset V} 
\hat{H}_X\;, 
\end{eqnarray} 
where the summation runs over the  subsets $X$ of $V$ with $\hat{H}_X$ being a self-adjoint operator that  is local on the Hilbert space  
 ${\cal H}_X:=  \otimes_{x\in X} {\cal H}_x$ , i.e. it acts 
 non-trivially on  the spins of $X$ while being the identity everywhere else.

Assume then that  Alice and Bob  control respectively two non-overlapping sections A and  B of the network ${\cal N}$, their distance being  $d(A,B)>0$. The model includes also a domain $C$ of ${\cal N}$ that  represents the
spins which are neither under Bob's nor Alice's control. 
The two parties agree about a protocol according to which Alice  signals  to Bob 
by locally perturbing the input state of the chain $\hat{\tau}_{ABC}$ via a set of local operations acting 
on the spins belonging to her domain $A$.  
Such actions will hence 
 propagate  according to the natural Hamiltonian (\ref{HAMILT}) of the network for some transferring time $t$
after which Bob will try to recover them via some proper local operations on the domain $B$. 
 The question we want to address is how much Bob will be able to discern about 
Alice's encoding action 
by performing arbitrary (local) operations on the output state~(\ref{DEFROB}). 
In the next section we shall approach this problem by generalizing the work of Ref.~\cite{C1} where, using 
the Lieb-Robinson (LR) inequality~\cite{LR,REVIEW}  an upper limit was set for the Holevo capacity $C_1$~\cite{REFCAP} attainable
 using a specific spin-network communication strategy (explicitly the model defined in Eq.~(\ref{DEFROBclass}) below). We remind that the LR is a universal bound on the 
 correlations that can be established between distant portions of the network due to  the dynamics induced
 by the system Hamiltonian $\hat{H}$ under minimal assumptions about the structure of involved couplings. 
 In particular, given 
 any two operators
   $\hat{A}$ and $\hat{B}$ that are local  on Alice's and Bob's subsets $A$ and $B$ respectively,
   the LR inequality imposes the constraint 
 \begin{eqnarray} 
\frac{\| [\hat{A}(t),\hat{B}]\|}{ \| \hat{A}\| \| \hat{B}\| }   &\leq& \epsilon_{AB}(t) \;, 
 \label{eq:MinNachBound}
\end{eqnarray}
where 
\begin{eqnarray} \label{OPNORM} \| \hat{\Theta} \|: = \max_{|\psi\rangle} \| \hat{\Theta} |\psi\rangle\|\;,\end{eqnarray}  represents the standard operator norm, and 
where given 
\begin{eqnarray} \label{UNITARY} 
\hat{U}(t) := \exp[ - i \hat{H} t ]\;,\end{eqnarray} 
 the unitary operator associated with the network Hamiltonian~(\ref{HAMILT}) ($\hbar=1$), 
 \begin{eqnarray} 
           \hat{A}(t) := \hat{U}^\dag(t) \hat{A} \hat{U}(t)\;,
           \end{eqnarray} 
is   the evolved counterpart of $\hat{A}$ in the Heisenberg representation. According to the LR analysis, the quantity $\epsilon_{AB}(t)$ appearing on the r.h.s. of (\ref{eq:MinNachBound}) exhibits an explicit dependence upon the coupling strengths but  is independent of the actual state of the network $\hat{\tau}_{ABC}$. Most importantly it depends upon $t$ via its absolute value $|t|$,
and  tends to zero when this parameter is small   and/or $d(A,B)$ is large enough, pointing out that 
 modifications on $A$ sites require a certain time to affect the sector $B$ when the two are disjoint.
In particular, as shown in Ref.~\cite{PRIMO}, for finite range Hamiltonians admitting $\bar{D}$ such that 
 $\hat{H}_X = 0$ whenever  $D(X) > \bar{D}$,
 we can express the LR quantity $\epsilon_{AB}(t)$ in the following compact form 
\begin{eqnarray} \label{DEFEPSILON}
  \epsilon_{AB}(t)  = 2|A| |B|     \left(\frac{2\, e\,\zeta\,\bar{D}\,|t|}{d(A,B)}\right)^{\tfrac{d(A,B)}{\bar{D}}}\;, 
\end{eqnarray} 
where $|X|$ is the total number of sites in  the domain $X\subset V$, and where $\zeta$ is a finite, positive constant characterizing the graph topology and the intensity of the couplings (but not on the size of the graph). 
If instead the Hamiltonian is explicitly of long-range couplings but sufficiently well behaved so that there exist
$\mu,s$ positive constants such that 
$\sup_{x \in V}  \sum\limits_{X\ni x}  \left|X\right|  \|\hat{H}_X\|  e^{\mu_2 D(X)}  \leq s$ (exponential decay), or 
 $\sup_{x \in V}  \sum\limits_{X\ni x}  \left|X\right|  \|\hat{H}_X\| [1+D(X)]^{\mu} \leq s$ (power-law decay), then 
Eq.~(\ref{DEFEPSILON})
gets replaced by 
\begin{equation}\label{eq: lambda boundnew2}
 \epsilon_{AB}(t) =
C |A| | B|  ( e^{v |t|} -1)  e^{-\mu d(A,B)}  \;, 
\end{equation}
in the first case, and by 
\begin{equation}\label{eq: lambda boundnew1}
 \epsilon_{AB}(t) =
C |A| | B|  \frac{ e^{v |t|} -1}{ (1+ d(A,B))^{\mu}} \;,
\end{equation}
in the second case, $v$ and $C$ being positive constants that again depend upon 
the metric of the network and on the Hamiltonian, but do not scale with the size of the model~\cite{exponential1,Clust Theo}.

\subsection{SNC channels} \label{sub2}

Without loss of generality we can describe  the perturbation induced by Alice on the network in an effort to communicate with Bob as a  Linear, Completely Positive,
Trace preserving (LCPT)~\cite{CHOI75,HOLEVOBOOK,WILDEBOOK,NC}  encoding map ${\cal E}_{QA}$  which  at time $t=0$ 
locally couples the portion $A$ of ${\cal N}$ with an external memory element $Q$ that stores the information 
she wants Bob to receive, see Fig.~\ref{fig:network}. Specifically, indicating with $\hat{\tau}_{ABC}$ the initial state of the network we have 
\begin{eqnarray} 
\hat{\rho}_{Q}\rightarrow 
{\cal E}_{QA}[\hat{\rho}_Q\otimes \hat{\tau}_{ABC}] := ({\cal E}_{QA}\otimes {\cal I}_{BC}) [\hat{\rho}_Q\otimes\hat{\tau}_{ABC}] \;, \label{MAPPoring} \end{eqnarray} 
 where ${\cal I}_{BC}$ represents the identity superoperator on the $BC$ domains.
Once introduced into the system,
the perturbation (\ref{MAPPoring})
propagates freely for a transferring  time $t$
 along the spin-network,  i.e.  \begin{equation}\label{eq: 1 Encoding symbols}
{\cal E}_{QA}[\hat{\rho}_Q\otimes \hat{\tau}_{ABC}] 
  \longrightarrow \hat{U}(t){\cal E}_{QA}[\hat{\rho}_Q\otimes \hat{\tau}_{ABC}] \hat{U}^{\dagger}(t)\;,
\end{equation}
with $\hat{U}(t)$ being the unitary transformation~(\ref{UNITARY}) defining the dynamics of ${\cal N}$.
Bob on his sites will have hence the possibility of  perceiving it as a modification of the reduced density matrix of the portion of spin-network he controls, i.e. 
\begin{eqnarray}  \label{DEFROB} 
\hat{\rho}_B(t)= \Phi[  \hat{\rho}_Q ] &:=& 
\mbox{Tr}_{QAC}\Big(\hat{U}(t){\cal E}_{QA}[\hat{\rho}_Q\otimes \hat{\tau}_{ABC}]\hat{U}^{\dagger}(t)\Big) \nonumber \\
&=& \mbox{Tr}_{AC}\Big( \hat{U}(t){\cal E}_{A}[ \hat{\tau}_{ABC}]\hat{U}^{\dagger}(t)\Big) \;, 
\end{eqnarray} 
where in the second line we  used the fact that $\hat{U}(t)$ does not operate on $Q$, to introduce the LCPT mapping locally acting on $A$ 
\begin{eqnarray} 
\hat{\tau}_{ABC}\rightarrow 
{\cal E}_A[\hat{\tau}_{ABC}] := \mbox{Tr}_Q\Big(
{\cal E}_{QA} [\hat{\rho}_{Q} \otimes \hat{\tau}_{ABC}]\Big) \;, \label{MAPP} \end{eqnarray} 
 that  depends on the selected message $\hat{\rho}_{Q}$ and encoding operation ${\cal E}_{QA}$.

Equation~(\ref{DEFROB})  defines 
the SNC channel $\Phi$ connecting Alice's quantum memory $Q$ to Bob's location. By construction it is explicitly LCPT and 
besides the properties of the network (namely its Hamiltonian $\hat{H}$ and its input state 
$\hat{\tau}_{ABC}$) and the propagation time $t$, 
it  explicitly depends upon Alice's choice of the encoding transformation ${\cal E}_{QA}$.
A trivial option is represented for instance by the case where ${\cal E}_{QA}$ is the identity 
mapping ${\cal I}_{QA}$: under this assumption no information is transferred from $Q$ either to the $A$ or to the $B$ portion of the
network, leading~(\ref{DEFROB})  to coincide with the 
  depolarizing map~\cite{DEP,NC} $\Phi_{DP}^{(0)}$ defined by the identity
\begin{eqnarray}  \label{DEPO} 
 \Phi_{DP}^{(0)}[ \hat{\rho}_Q] : = \hat{\rho}^{(0)}_B(t) \; \mbox{Tr} [ \hat{\rho}_Q] \;,
\end{eqnarray} 
where 
\begin{eqnarray}  \label{DEFROBunp} 
\hat{\rho}^{(0)}_B(t):= \mbox{Tr}_{AC}[ \hat{U}(t)\hat{\tau}_{ABC} \hat{U}^{\dagger}(t)]\;,
\end{eqnarray} 
is the state Bob  would have received if Alice decided not to perturb her spins at time $t=0$. 
Identifying instead ${\cal E}_{QA}$ with a control gate activated by different choices of $\hat{\rho}_{Q}$, we  can  force ${\cal E}_A$ to belong to a generic list 
$\{{\cal E}_{A}^{(\alpha)}\}_\alpha$ of possible operations, each associated with   
a classical symbol
labeled by the index $\alpha$. With this choice  the  scheme (\ref{DEFROB}) induces the mapping 
 \begin{eqnarray}  \label{DEFROBclass} 
\alpha\longrightarrow   \hat{\rho}_{B}^{(\alpha)}(t):= 
 \mbox{Tr}_{AC}\Big( \hat{U}(t) {\cal E}^{(\alpha)}_{A} [\hat{\tau}_{ABC}]\hat{U}^{\dagger}(t)\Big) \;, 
\end{eqnarray} 
that 
 corresponds  to the 
the signaling strategy  analysed in Ref.~\cite{C1} to allow the transferring of classical messages from $A$ to $B$.
 On the contrary, by identifying $Q$ with a memory element $Q_A$ that is  isomorphic with $A$ and taking ${\cal E}_{QA}$ to be a unitary swap gate,
Eq.~(\ref{MAPP}) reduces to 
 \begin{eqnarray}
\hat{\tau}_{ABC}\rightarrow \hat{\rho}_A \otimes \hat{\tau}_{BC} \;, \label{MAPP1} \end{eqnarray}  
with $\hat{\rho}_A$ being the isomorphic copy of $\hat{\rho}_{Q_A}$ on $A$ and 
$\hat{\tau}_{BC} := \mbox{Tr}_A[ \hat{\tau}_{ABC}]$ being the reduced state of the $BC$ domains
 obtained by tracing away $A$ from the input $\hat{\tau}_{ABC}$.
Accordingly, under this construction  the SNC channel~(\ref{DEFROB}) becomes  
 \begin{eqnarray}  \label{DEFROBswap} 
\Phi_{SW} [  \hat{\rho}_{Q_A} ] &=& 
 \mbox{Tr}_{AC}\Big( \hat{U}(t)[ \hat{\rho}_{A} \otimes \hat{\tau}_{BC}]\hat{U}^{\dagger}(t)\Big) \;, 
\end{eqnarray} 
which represents the swap-in/swap-out spin-network communication strategy extensively studied in the literature
(see e.g. 
Refs. \cite{BOSE,BOSE1,SPIN1,SPIN2,
SPIN3,SPIN4,SPIN5,SPIN6,DANIEL})
that, at least in principle, is capable
to convey both classical and quantum messages.

Of course, Eqs.~(\ref{DEPO}),~(\ref{DEFROBclass}), and (\ref{DEFROBswap}) are just three examples out of a large (possibly infinite) set of possible maps~(\ref{DEFROB}) that we can realize for fixed  $\hat{\tau}$, $\hat{H}$ and $t$,  by 
using different choices of the mapping ${\cal E}_{QA}$. 
Determining  what is the optimal option in terms of communication efficiency 
is a rather complex problem which arguably depends upon the property of the network, the value of 
transferring time $t$, the relative distance of the  locations $A$ and $B$, 
as well as upon the kind of messages (classical, private classical, quantum, etc.) one wishes to transfer. 
Our aim is to show that however, irrespectively of the freedom to select the encoding ${\cal E}_{QA}$,  the LR inequality 
(\ref{eq:MinNachBound}) poses a fundamental limitation on the resulting 
 communication efficiency.

\section{Distance of the received message from the non-signaling state} \label{Sec:nonsig}

To determine the amount of information that can be effectively retrieved by Bob at the end of the
transmission~(\ref{DEFROB}) associated with an 
arbitrary coding strategy ${\cal E}_{QA}$, 
we have to compute the distance between the
 SNC channel $\Phi$ and the depolarizing channel $\Phi_{DP}^{(0)}$ of Eq.~(\ref{DEPO})
associated with the non-signaling protocol.
Specifically in Sec.~\ref{tracedist} we first analyze the induced trace-norm distance~\cite{WatrousBOOK}  between
$\Phi$ and $\Phi_{DP}^{(0)}$ showing that irrespectively of the choice of ${\cal E}_{QA}$ we get the inequality 
\begin{eqnarray} 
\| \Phi - \Phi_{DP}^{(0)} \|_1 
\leq   M_A^2 \;  {\epsilon}_{AB}(t)\;, \label{TRACEB} 
\end{eqnarray} 
where $M_A$ is the dimension of the 
Hilbert space associated with the spins of the domain $A$ under Alice's control and where
$\epsilon_{AB}(t)$ is the LR quantity appearing on the r.h.s. of  Eq.~(\ref{eq:MinNachBound}). 
Equation~(\ref{TRACEB}) is a clear indication that for small enough values of $t$ and/or large enough values of $d(A,B)$,
the spin-network channel performances are close to the non-signaling regime,
 irrespectively of the initial state $\hat{\tau}_{ABC}$ of the
network and from the encoding procedure ${\cal E}_{QA}$ selected by Alice. 
In particular  from Eq.~(75) of  Ref.~\cite{SHIROKOV}  
it is possible to use
 Eq.~(\ref{TRACEB}) to bound the value of 
the Holevo capacity~\cite{CAP1,CAP2} associated with $\Phi$ as 
\begin{eqnarray} \label{C1BOUND} 
 C_1(\Phi) &\leq& \frac{M_A^2 \;  {\epsilon}_{AB}(t)}{2}  \log_2 M_B + g( \tfrac{M_A^2 \;  {\epsilon}_{AB}(t)}{2})\;, 
 \end{eqnarray} 
 where we exploited the fact that $C_1(\Phi_{DP}^{(0)})$ is trivially null (no information being transferred via 
 the depolarizing map) and where
  $g(x)$ is a function that tends to zero as $x\rightarrow 0$, defined by the identities 
\begin{eqnarray} 
g(x) &:=& (1+x) H_2 (x/(1+x)) \;, \\ H_2(y) &:=& - y \log_2y - (1-y) \log_2(1-y) \;.
\end{eqnarray} 
Equation (\ref{C1BOUND}) generalizes an analogous result obtained in Ref.~\cite{C1}
in the special case of the classical-to-quantum encoding strategy~(\ref{DEFROBclass}).
Extending this to  all possible encodings and to the full set of communication capacities~\cite{REFCAP,HOLEVOBOOK,WILDEBOOK} (i.e. 
the classical capacity $C(\Phi)$~\cite{CAP1,CAP2}, the private capacity $C_P(\Phi)$~\cite{CAP3},
the quantum capacity $Q(\Phi)$~\cite{CAP3,CAP4,CAP5},  and the entanglement assisted capacity $C_E(\Phi)$~\cite{CE1,CE2} of the map $\Phi$),  
requires however a little more effort. For this purpose 
in Sec.~\ref{secdiamond} we 
focus on the 
diamond distance~\cite{DIAMOND,DIAMOND1}  between $\Phi$ and 
a slightly different version of the depolarizing channel $\Phi_{DP}^{(0)}$, namely the channel 
 \begin{eqnarray}  \label{DEPOnew} 
 \Phi^{(1)}_{DP}[ \hat{\rho}_Q] : = \hat{\rho}^{(1)}_B(t) \; \mbox{Tr} [ \hat{\rho}_Q] \;,
\end{eqnarray} 
 obtained by 
replacing in Eq.~(\ref{DEPO}) the state 
$\hat{\rho}^{(0)}_B(t)$ of (\ref{DEFROBunp}) with the density matrix 
\begin{eqnarray}  \label{DEFROBunp1} 
\hat{\rho}^{(1)}_B(t):= \mbox{Tr}_{AC}[ \hat{U}(t)(\hat{\tau}_A\otimes \hat{\tau}_{BC} )\hat{U}^{\dagger}(t)]\;,
\end{eqnarray} 
with $\hat{\tau}_A:= \mbox{Tr}_{BC}[\hat{\tau}_{ABC}]$ and $\hat{\tau}_{BC}:=\mbox{Tr}_A [\hat{\tau}_{ABC}]$ the reduced density matrices of the sectors ($A$ and $BC$ respectively) of the input state of the network $\hat{\tau}_{ABC}$.
According to our analysis we shall see that the following inequality holds
  \begin{eqnarray}  \label{defdiamINEQ} 
\| \Phi - \Phi^{(1)}_{DP} \|_{\Diamond} \leq    M \;  {\epsilon}_{AB}(t)\;,
\end{eqnarray}
where again ${\epsilon}_{AB}(t)$ is the  LR quantity and where $M$ 
is upper bounded by $2 M_A^4$, specifically
 \begin{eqnarray}  \label{MMM144} 
 M&:=& 2 \;  \mbox{min}\{M_A^4 ,M_A^3 M_B M_C\}\;.\end{eqnarray}
Notice that as for Eq.~(\ref{TRACEB}), 
the r.h.s. of this inequality involves only quantities that ultimately just depend upon properties of the spin-network: specifically the distance of the sectors $A$ and $B$, the number of spins they contain, the transferring time $t$, the dimension of the Hilbert space of $A$.
From the results of Leung and Smith~\cite{LEUNG} and the subsequent
improvement by  Shirokov~\cite{SHIROKOV} we can now turn Eq.~(\ref{defdiamINEQ}) into a
 bound for the communication capacities~\cite{REFCAP,HOLEVOBOOK,WILDEBOOK}
 of the map $\Phi$ in terms of the corresponding ones associated with the depolarizing map  $\Phi^{(1)}_{DP}$.
Explicitly, observing that by
definition we have \begin{eqnarray} 
C_1(\Phi^{(1)}_{DP})&=&C(\Phi^{(1)}_{DP}) =0\;, \nonumber \\
  C_P(\Phi^{(1)}_{DP})  &=& Q(\Phi^{(1)}_{DP}) =0 \;, \nonumber \\
C_E(\Phi^{(1)}_{DP}) &=&0\;,\end{eqnarray} 
 equations~(81) and (82) of Ref.~\cite{SHIROKOV} lead us to 
\begin{equation}
Q(\Phi), C(\Phi) \leq M \;  {\epsilon}_{AB}(t)  \log_2 M_B + g( \tfrac{M\;  {\epsilon}_{AB}(t)}{2})\;, 
\label{QCCP} 
\end{equation} 
while Eq.~(76) of Ref.~\cite{SHIROKOV} to
\begin{equation} \label{CEB} 
C_E(\Phi) \leq  M \;  {\epsilon}_{AB}(t)  \log_2 M' + g(\tfrac{M \;  {\epsilon}_{AB}(t)}{2}) \;,
\end{equation} 
 where 
  $M'$ is the minimum between the dimensions of $A$ and $B$, i.e. 
  \begin{eqnarray} M': = \min\{ M_A, M_B\}\;.\end{eqnarray}
As a matter of fact the last of the inequalities presented above happens to be the strongest of all:
indeed due to the natural ordering among the capacities \cite{CapHierarchy}
\begin{eqnarray} 
 C_P(\Phi) \leq C(\Phi) \leq C_E(\Phi) \;,\qquad 
 Q(\Phi) \leq C_E(\Phi)/2\;, 
\end{eqnarray} 
our final bounds read 
\begin{equation}
C_P(\Phi), C(\Phi), C_E(\Phi) \leq M \;  {\epsilon}_{AB}(t)  \log_2 M' +g( \tfrac{M \;  {\epsilon}_{AB}(t)}{2})\;, 
\end{equation} 
\begin{equation} 
Q(\Phi) \leq \frac{M \;  {\epsilon}_{AB}(t)}{2}   \log_2 M' +\frac{1}{2} g( \tfrac{M \;  {\epsilon}_{AB}(t)}{2}) \;.
\end{equation} 

 \subsection{Induced trace-norm distance} \label{tracedist}

The induced trace distance between 
$\Phi$ of Eq.~(\ref{DEFROB}) and the depolarizing channel $\Phi_{DP}^{(0)}$ of Eq.~(\ref{DEPO})
related to the non-signaling protocol is defined as 
\begin{eqnarray} \| \Phi - \Phi_{DP}^{(0)} \|_1\label{TRACEDIST} 
:= 2 \max_{\hat{\rho}_Q} D(\Phi(\hat{\rho}_Q),\Phi_{DP}^{(0)}(\hat{\rho}_Q) )\;,
\end{eqnarray} 
where 
the maximum is taken over the whole set of possible input states $\hat{\rho}_Q$ of the memory
$Q$, and $D(\Phi(\hat{\rho}_Q),\Phi_{DP}^{(0)}(\hat{\rho}_Q))$ is the trace-distance~\cite{NC} between
the corresponding output configurations $\hat{\rho}_B(t)$ and $\hat{\rho}^{(0)}_B(t)$ of $\Phi$ and $\Phi_{DP}^{(0)}$. 
According to the Helstrom theorem~\cite{HOLEVOBOOK,WILDEBOOK}, $D(\Phi(\hat{\rho}_Q),\Phi_{DP}^{(0)}(\hat{\rho}_Q))$ gauges the minimum error probability that one can get trying to discriminate $\Phi(\hat{\rho}_Q)$ from $\Phi_{DP}^{(0)}(\hat{\rho}_Q))$, in particular it writes
\begin{eqnarray} 
D(\Phi(\hat{\rho}_Q),\Phi_{DP}^{(0)}(\hat{\rho}_Q))&=&D(\hat{\rho}_B(t),\hat{\rho}^{(0)}_B(t) )\nonumber \\
&: =& \frac{1}{2} \| \hat{\rho}_B(t) -
\hat{\rho}^{(0)}_B(t)\|_1\;,  \label{TRACED} 
\end{eqnarray} 
with $\| \hat{X} \|_1:= \mbox{Tr}[ \sqrt{\hat{X}^\dag \hat{X}}]$ being the trace-norm of the operator $
\hat{X}$, not to be confused with the operator norm introduced in Eq.~(\ref{OPNORM}).
A useful way to express~(\ref{TRACED}) is 
\begin{equation} 
D(\hat{\rho}_B(t),\hat{\rho}^{(0)}_B(t) )
=  \max_{\hat{\Theta}_B} \left| \mathrm{Tr}_B\left[\hat{\Theta}_B (\hat{\rho}_B(t) -
\hat{\rho}^{(0)}_B(t)) \right] \right|\;,
\end{equation} 
where the maximum can be taken   either over the set of positive operators $\hat{\openone}_B \geq \hat{\Theta}_B\geq 0$, or, equivalently, on the set of operators 
$\hat{\Theta}_B = \hat{V}_B/2$ with $\hat{V}_B$ being a unitary operator acting locally on the spins of the domain $B$ (in what follows we'll find more convenient the latter option). 
Introducing 
the operator $\hat{\Theta}_B(t):=\hat{U}^\dagger(t)\hat{\Theta}_B\hat{U}(t)$ and using 
 Eqs.~(\ref{eq: 1 Encoding symbols}), (\ref{DEFROB}), and (\ref{DEFROBunp}) we can then write 
\begin{eqnarray}
D(\hat{\rho}_B(t),\hat{\rho}^{(0)}_B(t) )\nonumber 
=\max_{\hat{\Theta}_B}\left| \mathrm{Tr} \left[\hat{\Theta}_B(t) \; 
({\cal E}_A[ \hat{\tau}_{ABC}] -  \hat{\tau}_{ABC}) \right]\right|&&\\
\quad = \max_{\hat{\Theta}_B} \left| \sum_{k=1}^K \mathrm{Tr}\left[ \hat{M}_k^\dagger \hat{\Theta}_B(t) \hat{M}_k \hat{\tau}_{ABC}- \hat{\Theta}_B(t)  \hat{M}_k^\dagger \hat{M}_k\hat{\tau}_{ABC} \right] \right|&&\nonumber  \\
\qquad = \max_{\hat{\Theta}_B} \left| \sum_{k=1}^K \mathrm{Tr}\left[ [ \hat{M}_k^\dagger, \hat{\Theta}_B(t)]\;  \hat{M}_k \;  \hat{\tau}_{ABC} \right] \right|\;, && \nonumber 
\end{eqnarray}
where  $\{ \hat{M}_{k}; {k=1,\cdots, K}\}$ 
are a Kraus set of local operators on $A$ which represents the action of  the LCPT map ${\cal E}_A$, i.e. 
\begin{equation} 
\label{KRAUS} 
{\cal E}_A[ \cdots] =\sum_{k=1}^K \hat{M}_{k} [ \cdots] \hat{M}_k^\dagger, \qquad 
\sum_{k=1}^K \hat{M}_k^\dagger\hat{M}_{k}   = \hat{\openone}\;.
\end{equation}
Now bounding the expectation value of $[\hat{M}_k^\dagger, \hat{\Theta}_B(t)]  \hat{M}_k $ over $\hat{\tau}_{ABC}$ with the associated operator norm (\ref{OPNORM}), exploiting the triangular inequality we obtain 
\begin{eqnarray}
D(\hat{\rho}_B(t),\hat{\rho}^{(0)}_B(t) ) 
 \leq \max_{\hat{\Theta}_B} \sum_k^K\| [\hat{M}_k^\dagger,\hat{\Theta}_B(t)]\| \|  \hat{M}_k\| \;,
 \label{DISE}
 \end{eqnarray} 
Observe that by unitary equivalence of the norm we have 
$\| [\hat{M}_k^\dagger,\hat{\Theta}_B(t)]\| = \| [ \hat{M}_k^\dagger(-t),\hat{\Theta}_B]\|$
where now $\hat{M}_k^\dagger(t) = \hat{U}^\dag(t) \hat{M}_k^\dagger \hat{U}(t)$ is the time evolved
version of the local operator $\hat{M}_k^\dagger$ of $A$ under the action of the network Hamiltonian.
Accordingly we can use (\ref{eq:MinNachBound}) and (\ref{DEFEPSILON}) to write 
\begin{eqnarray}
\| [\hat{M}_k^\dagger,\hat{\Theta}_B(t)]\| 
\leq  \| \hat{M}_k^\dagger\| \| \hat{\Theta}_B\|   \epsilon_{AB}(t)  
\leq   \epsilon_{AB}(t) /2\;, 
\end{eqnarray} 
where we used the fact that 
\begin{eqnarray} \| \hat{M}_k^\dag\| = \| \hat{M}_k\|=\sqrt{ \| \hat{M}_k^\dag \hat{M}_k\| } \leq 1\;,\end{eqnarray}  due to the normalization condition of the Kraus elements, and 
$ \| \hat{\Theta}_B\| = \| \hat{V}_B/2 \| \leq 1/2$.
Replacing this into the bound on  $D(\hat{\rho}_B(t),\hat{\rho}^{(0)}_B(t) )$ we hence can write 
\begin{equation}
D(\hat{\rho}_B(t),\hat{\rho}^{(0)}_B(t) )
\leq   (K /2)  \;\; \epsilon_{AB}(t) \;, 
 \label{BOUND0} 
\end{equation} 
with the r.h.s. that  depends upon the specific choice of the encoding channel ${\cal E}_A$ only via the total number $K$ of Kraus elements
that enter the decomposition (\ref{KRAUS}). 
In case we restrict Alice to adopt only unitary encodings, this yields $K=1$. Alternatively, if we
allow for arbitrary LCPT operations ${\cal E}_A$ on $A$, i.e. arbitrary LCPT operations ${\cal E}_{QA}$ on $Q$ and $A$,  an universal bound can be established by reminding that, irrespectively of the choice of ${\cal E}_A$ 
it is always possible to have a Kraus set with at most $K = M_A^2$ \cite{CHOI75}.
This leads to 
\begin{eqnarray} \label{BOUNDDDD1} 
&&D(\hat{\rho}_B(t),\hat{\rho}^{(0)}_B(t) )
\leq( M_A^2/2)\;  {{\epsilon}_{AB}(t)}  \;, 
\end{eqnarray} 
and hence to Eq.~(\ref{TRACEB}) via Eq.~(\ref{TRACEDIST}) exploiting   the fact that 
 the r.h.s. of  Eq.~(\ref{BOUNDDDD1}) holds true for all possible choices of the input $\hat{\rho}_Q$.

\subsection{Diamond norm distance} \label{secdiamond}

 The diamond-distance~\cite{DIAMOND,DIAMOND1} between two channels $\Phi$ and $\Phi'$ connecting $Q$ to $B$ 
 is defined as 
  \begin{equation}  \label{defdiam} 
  \| \Phi - \Phi'  \|_{\Diamond} = 
   \max_{|\psi\rangle_{QQ'}} \| (\Phi\otimes {\cal I} - 
  \Phi' \otimes {\cal I})(|\psi\rangle_{QQ'}\langle \psi|) \|_1 \;, 
\end{equation}
where the maximization now is performed for   extensions
$\Phi\otimes {\cal I}$ and $\Phi'\otimes {\cal I}$  of the original channels involving purifications $|\psi\rangle_{QQ'}$ of the possible 
inputs of $Q$ constructed on 
an ancillary system $Q'$  that is isomorphic to $Q$. 
A naive way to bound this quantity would be given by using the natural ordering with the
induced trace-norm distance (see  Appendix~\ref{APPB}), according to which 
one has  
 \begin{eqnarray} 
 \| \Phi - \Phi' \|_{1} \leq \| \Phi - \Phi' \|_{\Diamond}  \leq  2 M_Q \| \Phi - \Phi' \|_{1} \;,
 \label{diamondnatural} 
\end{eqnarray}
with $M_Q$ being the dimension of Alice's memory $Q$. Applying this to the maps 
$\Phi$, associated with a generic encoding ${\cal E}_{QA}$, and to the depolarizing channel $\Phi_{DP}^{(0)}$ of Eq.~(\ref{DEPO}) 
yields 
 \begin{equation} 
 \| \Phi - \Phi_{DP}^{(0)} \|_{\Diamond}  \leq  2 M_Q \| \Phi - \Phi_{DP}^{(0)} \|_{1}
 \leq 
 2 M_Q M_A^2 \epsilon_{AB}(t) \;,
 \label{diamondnatural111} 
\end{equation} 
 where in writing the last term we invoked the bound (\ref{TRACEB}). 
In many cases of physical interest  where $M_Q$ is directly linked to the dimensionality of $A$, 
Eq.~(\ref{diamondnatural111}) is sufficiently strong for our purposes. For instance this happens 
for the swap-in/swap-out coding map $\Phi_{SW}$ of Eq.~(\ref{DEFROBswap}),  
where by construction the memory element is isomorphic to $A$, i.e. $M_{Q_A}=M_A$. Accordingly,
in this case Eq.~(\ref{diamondnatural111}) leads  to
 \begin{eqnarray} 
\| \Phi_{SW}- \Phi_{DP}^{(0)} \|_{\Diamond}  \leq  2 M_A^3 \epsilon_{AB}(t)  \;,
 \label{diamondnatural1} 
\end{eqnarray}
which can be used to replace~(\ref{defdiamINEQ}) in our study of the channel capacities reported at the beginning of Sec.~\ref{Sec:nonsig}. 
 For a generic choice of ${\cal E}_{QA}$ however, the presence of $M_Q$ on the
 r.h.s. of Eq.~(\ref{diamondnatural111}) poses a severe limitation to this inequality as the dimension of $Q$ 
 is not a property of the spin-network model and can in principle assume unbounded values. 
To deal with this problem we now consider the diamond norm
 \begin{equation}  \label{defdiameew} 
  \| \Phi -  \Phi_{DP}^{(1)}  \|_{\Diamond} = 
   \max_{|\psi\rangle_{QQ'}} \| (\Phi\otimes {\cal I} - 
  \Phi_{DP}^{(1)}  \otimes {\cal I})(|\psi\rangle_{QQ'}\langle \psi|) \|_1 \;, 
\end{equation}
  between the map $\Phi$ associated
with the encoding operation ${\cal E}_{QA}$ and the depolarizing map $\Phi^{(1)}_{DP}$ defined in Eq.~(\ref{DEPOnew}).
Notice that the actions of $\Phi$ and $\Phi^{(1)}_{DP}$ can be expressed as a concatenation of two processes, i.e.   
 \begin{eqnarray}
\Phi[\cdots] &=& \Psi \circ {\cal E}[\cdots] \;, \label{dd1} \\ 
\Phi^{(1)}_{DP}[\cdots]&=& \Psi^{(1)}_{DP} \circ {\cal E}[\cdots] \;,  \label{dd2} 
 \end{eqnarray}
 where 
  \begin{eqnarray} \label{MAPEEEE} 
 {\cal E}[\cdots]:= \mbox{Tr}_Q [{\cal E}_{QA}[\cdots\otimes \hat{\tau}_{ABC}]]\;, 
 \end{eqnarray} 
 is a LCPT channel from $Q$ to $ABC$ and where 
 \begin{eqnarray}
 \Psi[ \cdots ] &:=& \mbox{Tr}_{AC}\Big[ \hat{U}(t)[ \cdots ]\hat{U}^{\dagger}(t)\Big]\;, \\
  \Psi_{DP}^{(1)}[ \cdots ]&:=& \mbox{Tr}_{AC}\Big[ \hat{U}(t)\left( \hat{\tau}_A \otimes \mbox{Tr}_{A}[ \cdots ]\right)\hat{U}^{\dagger}(t)\Big]\;, \nonumber\\
 \end{eqnarray}
 are instead LCPT transformations operating from $ABC$ to $B$ which do not depend upon the special choice of ${\cal E}_{QA}$. 
 
Consider  first the case where the input state  $\hat{\tau}_{ABC}$ of the network ${\cal N}$ is a pure vector $\ket{\tau}_{ABC}$.
For a generic choice of the pure states  $|\psi\rangle_{QQ'}$  of $QQ'$ entering the maximization~(\ref{defdiameew}),  
we have that  globally the $QQ'ABC$ system is described by the product vector 
$\ket{\psi,\tau}_{QQ'ABC}:=\ket{\psi}_{QQ'}\ket{\tau}_{ABC}$, 
which, after a Schmidt decomposition of $\ket{\psi}_{QQ'}$ and $\ket{\tau}_{ABC}$ along the partitions $Q,Q'$ and $A, BC$ respectively, can be written as
\begin{eqnarray}
&&\ket{\psi,\tau}_{QQ'ABC}= \sum_{i=1}^{r} \sum_{j=1}^{s}\sqrt{\alpha_i \beta_j}\ket{\psi_i,\psi_i,\tau_j,\tau_j}_{Q'QABC}\;, \nonumber   \\  
&&\ket{\psi_i,\psi_i,\tau_j,\tau_j}_{Q'QABC}:=\ket{\psi_i}_{Q'} \ket{\psi_i}_{Q}\ket{\tau_j}_A \ket{\tau_j}_{BC},  \nonumber \\ \label{defeee} 
\end{eqnarray}  
with $r\leq M_Q$ and $s\leq \mbox{min}\{M_A,M_B M_C\}$ with $\ket{\psi_i}_{Q/Q'}$ and $\ket{\tau_j}_{A/BC}$ forming an orthogonal set of pure states of their respective systems.
Completing  hence $\ket{\psi_i}_{Q}$ to a basis of $Q$, we then define the vectors
\begin{equation}
\ket{\tilde\lambda_{\ell,q}}_{Q' ABC}:=\sum_{i=1}^{r} \sum_{j=1}^{s} \sqrt{\alpha_i \beta_j}\ket{\psi_i}_{Q'} 
|\chi_{\ell,q,i,j}\rangle_A
 \ket{\tau_j}_{BC}\;,  \label{DEFEEE1} 
\end{equation}
where 
\begin{eqnarray}
\label{ddfd} 
|\chi_{\ell,q,i,j}\rangle_A:= {_{Q}\bra{\psi_q}} \hat{N}_\ell \ket{\psi_i,\tau_j}_{QA}\;, \label{DEFW} \end{eqnarray} 
and where $\hat{N}_{\ell}$ are the Kraus operators associated with the channel ${\cal E}_{QA}$ 
 \begin{equation} 
\label{KRAUS_QA} 
{\cal E}_{QA}[ \cdots] =\sum_{\ell=1}^L \hat{N}_{\ell} [ \cdots] \hat{N}_\ell^\dagger, \qquad 
\sum_{\ell=1}^L \hat{N}_\ell^\dagger\hat{N}_{\ell}   = \hat{\openone}\;,
\end{equation}
with $L$ which can be always chosen to be smaller than $M^2_QM_A^2$.
Upon normalization Eq.~(\ref{DEFEEE1}) gives the pure states 
\begin{equation}\label{PUREST} 
\ket{\lambda_{\ell,q}}_{Q' ABC}:=|\tilde\lambda_{\ell,q}\rangle_{Q' ABC}/g_{\ell,q}\;, 
\end{equation} 
the norms $g_{\ell,q}:=\| |{\tilde\lambda_{\ell,q}}\rangle_{Q' ABC}\|$ 
satisfying the constraint 
\begin{equation}\label{CONVE} 
\sum_{\ell =1}^{L}\sum_{q=1}^{M_Q} g^2_{\ell,q}=1.
\end{equation}
Notice that since terms (\ref{ddfd}) are elements 
of the Hilbert space of ${\cal H}_A$,  it follows that for each given $q$ 
 and $\ell$, when varying indexes  $i$, $j$, vectors $|\chi_{\ell,q,i,j}\rangle_A \ket{\tau_j}_{BC}$ span a space of dimension not larger than 
 \begin{eqnarray}  \label{MMM1} 
 M_*&:=& M_A \times \mbox{min}\{M_A ,M_B M_C\}\nonumber \\
 &=&  \mbox{min}\{M_A^2 ,M_A M_B M_C\}\;.\end{eqnarray}
Accordingly this number also bounds the maximum number of non-zero terms entering  the Schmidt decomposition of $\ket{\lambda_{\ell,q}}_{Q' ABC}$ along the partition $Q', ABC$, i.e. 
\begin{eqnarray} \label{RANK} 
\ket{\lambda_{\ell,q}} = \sum_{m=1}^{M_*} \sqrt{\gamma_m} |m\rangle_{Q'} |m\rangle_{ABC} \;,
\end{eqnarray} 
for a proper choice of orthogonal sets of vectors  $|m\rangle_{Q'}$ and $|m\rangle_{ABC}$. 
Exploiting the above identities  the state of $Q'BC$ after the encoding stage through the mapping Eq.~(\ref{MAPEEEE})
can be casted in the following form 
\begin{equation} \label{DD112} 
{\cal E}\otimes {\cal I}\Big[\ket{\psi,\tau}\bra{\psi,\tau}\Big]
 =\sum_{\ell=1}^{L}\sum_{q=1}^{M_Q} g^2_{\ell,q} \ket{\lambda_{\ell,q}} \bra{\lambda_{\ell,q}}\;,
 \end{equation}
 where for ease of notation we set $\ket{\psi,\tau}:=\ket{\psi,\tau}_{QQ'ABC}$ and
  $\ket{\lambda_{\ell,q}}:= \ket{\lambda_{\ell,q}}_{Q'ABC}$. 
From (\ref{dd1}) and (\ref{dd2}) we hence get 
 \begin{eqnarray} \label{UESTA} 
&&\|(\Phi\otimes {\cal I} - \Phi^{(1)}_{DP}\otimes {\cal I})[\ket{\psi}_{QQ'}\bra{\psi}]\|_1  \\
 &&\qquad =  \| \sum_{\ell=1}^{L}\sum_{q=1}^{M_Q} g^2_{\ell,q} (\Psi \otimes {\cal I} - \Psi_{DP}^{(1)}\otimes {\cal I})[\ket{\lambda_{\ell,q}}\bra{\lambda_{\ell,q}}] \|_{1}\nonumber\\
 &&\qquad \leq  \sum_{\ell=1}^{L}\sum_{q=1}^{M_Q} g^2_{\ell,q}\|  (\Psi \otimes {\cal I} - \Psi_{DP}^{(1)}\otimes {\cal I})[\ket{\lambda_{\ell,q}}\bra{\lambda_{\ell,q}}] \|_{1}, \nonumber  \end{eqnarray}
the last inequality deriving from Eq.~(\ref{CONVE}) by convexity of the trace-norm. 
Remember now that each one of the vectors $\ket{\lambda_{\ell,q}}$ has Schmidt rank smaller than $M_*$ as indicated in Eq.~(\ref{RANK}).
Therefore, being the following steps identical to those in Appendix \ref{APPB} we get 
\begin{eqnarray}
&&\|  (\Psi\otimes {\cal I} - \Psi_{DP}^{(1)}\otimes {\cal I})[\ket{\lambda_{\ell,q}}\bra{\lambda_{\ell,q}}] \|_{1}\label{Fdet}  \\
&&\quad \leq  \sum_{m,m'=1}^{M_*} \sqrt{\gamma_m \gamma_{m'} } \Big\|  (\Psi - \Psi_{DP}^{(1)})[\ket{m}\bra{m'}]\otimes \ket{m}\bra{m'} \Big\|_{1} \nonumber \\ 
&&\quad = \sum_{m,m'=1}^{M_*} \sqrt{\gamma_m \gamma_{m'}}  \|  (\Psi - \Psi_{DP}^{(1)})[\ket{m}\bra{m'}]\|_{1} \nonumber \\ 
&&\quad \leq  2 {M_*}   \|  \Psi - \Psi_{DP}^{(1)}\|_{1}\;,
\end{eqnarray}
with $\|  \Psi - \Psi_{DP}^{(1)}\|_{1}$ being the induced trace-distance between $\Psi$ and $\Psi_{DP}^{(1)}$, i.e. the quantity 
\begin{eqnarray} \| \Psi - \Psi_{DP}^{(1)}  \|_1\label{TRACEDIST11} 
:=  \max_{\hat{\tau}'_{ABC}} \| (\Psi-\Psi_{DP}^{(1)})[\hat{\tau}'_{ABC}] \|_1\;.
\end{eqnarray} 
A crucial observation now is that, indicating with $Q_A$ Alice's memory which is isometric to $A$,  
 for all $\hat{\tau}'_{ABC}$ we can write 
\begin{eqnarray} 
\Psi[\hat{\tau}'_{ABC}]&=&\mbox{Tr}_{AC}[ \hat{U}(t) \hat{\tau}'_{ABC} \hat{U}^\dag(t)]= \Phi'_{DP}[\hat{\tau}_{Q_A}] \;, \nonumber \\ 
\Psi_{DP}^{(1)}[\hat{\tau}'_{ABC}]&=& \Phi'_{SW}[  \hat{\tau}_{Q_A} ] \;, 
\end{eqnarray} 
where
 $\hat{\tau}_{Q_A}$ represents the copy of $\hat{\tau}_A$ on $Q_A$, while $\Phi'_{DP}$ and $\Phi'_{SW}$ are respectively the non-signaling and the swap-in/swap-out
 channels associated with the input state  $\hat{\tau}'_{ABC}$  of the network. 
 Hence invoking (\ref{TRACEB}) we can write 
\begin{eqnarray}
\| (\Psi-\Psi_{DP}^{(1)}) [\hat{\tau}'_{ABC}] \|_1 
&=& \| (\Phi'_{DP} -\Phi')[\hat{\tau}_{Q_A}] \|_1  \\
&\leq& \| \Phi'_{DP} -\Phi' \|_1 \leq M_A^2 \;  {\epsilon}_{AB}(t)\;,  \nonumber 
\end{eqnarray} 
which, by reminding that ${\epsilon}_{AB}(t)$ does not depend upon the initial state of the spin-network, gives
\begin{eqnarray} \| \Psi - \Psi_{DP}^{(1)}  \|_1\label{TRACEDIST112} 
\leq  M_A^2 \;  {\epsilon}_{AB}(t)\;.
\end{eqnarray} 
Accordingly from Eq.~(\ref{TRACEDIST11})  and (\ref{UESTA}) 
we have \begin{equation}
\|(\Phi\otimes {\cal I} - \Phi^{(1)}_{DP}\otimes {\cal I})[\ket{\psi}_{QQ'}\bra{\psi}]\|_1 
\leq  2 M_* M_A^2 \;  {\epsilon}_{AB}(t)\;,
\end{equation} 
for all $\ket{\psi}_{QQ'}$, which replaced into Eq.~(\ref{defdiameew}) leads to 
 \begin{equation}  \label{defdiam11134} 
  \| \Phi - \Phi^{(1)}_{DP} \|_{\Diamond} \leq 2 M_*  M_A^2 \;  {\epsilon}_{AB}(t)\;,
\end{equation}
hence proving Eq.~(\ref{defdiamINEQ}). 

The above argument can be also used to deal with the   case where the initial state of the network $\hat{\tau}_{ABC}$ is not pure.
Indeed, by writing it as a convex sum over a set of pure states
\begin{eqnarray}
\label{conv111} 
\hat{\tau}_{ABC}= \sum_i p_i \ket{\tau_i}_{ABC}\bra{\tau_i}\;,\end{eqnarray} 
equation~(\ref{DD112}) gets replaced by  
\begin{equation}
{\cal E}\otimes {\cal I}\Big[\ket{\psi}\bra{\psi} \otimes  \hat{\tau}_{ABC} \Big]
 =\sum_i \sum_{\ell=1}^{L}\sum_{q=1}^{M_Q} p_i \left( g^{(i)}_{\ell,q} \right)^2 \ket{\lambda^{(i)}_{\ell,q}} \bra{\lambda_{\ell,q}^{(i)}}\;,
 \end{equation}
with $g^{(i)}_{\ell,q}$ and $\ket{\lambda^{(i)}_{\ell,q}}$ being associated with the $i$-th pure vector $\ket{\tau_i}_{ABC}$ entering Eq.~(\ref{conv111}) via the 
construction detailed in Eqs.~(\ref{defeee}-\ref{PUREST}). Consequently we can still invoke convexity to arrive at 
\begin{eqnarray} \label{UESTA1} 
&&\|(\Phi\otimes {\cal I} - \Phi^{(1)}_{DP}\otimes {\cal I})[\ket{\psi}_{QQ'}\bra{\psi}]\|_1  \\
 && \leq \sum_i  \sum_{\ell=1}^{L}\sum_{q=1}^{M_Q} p_i \left( g^{(i)}_{\ell,q} \right)^2 \|  (\Psi \otimes {\cal I} - \Psi_{DP}^{(1)}\otimes {\cal I})[
|{\lambda^{(i)}_{\ell,q}} \rangle\langle{\lambda_{\ell,q}^{(i)}}|] \|_{1}, \nonumber  \end{eqnarray}
that formally replaces (\ref{UESTA}). From here we can exploit the same steps reported in Eqs.~(\ref{Fdet}-\ref{defdiam11134}). 

\section{Conclusions} \label{Sec:conc} 

We propose a study of a broad set of information capacities associated with spin-networks employed as means of communication. In our analysis we considered as a quantum channel $\Phi$ a generic spin-network in a generic initial state equipped with an encoding represented by a local LCTP map, which results to be  more general with respect to specific solutions adopted previously in the literature.   Here we made use of the tools offered by the diamond norm and we exploited established results such as the Lieb-Robinson bound~\cite{LR}, which describes how correlations spread in spin systems, and Fannes inequality~\cite{FANNES}, which states continuity properties of the Von Neumann entropy. We were able in such a way to upper bound the whole set of quantum capacities of the map $\Phi$. Possible extensions of our work should include the presence of memory effects~\cite{MEMORY}  in the information transferring
which may arise when allowing Alice to perform sequences of encoding operations during the time
it takes for one of them to reach Bob's location.\\
 
The Authors would like to thank M. Shirokov, F. Verstraete, L. Lami and A. Winter for useful comments and suggestions.

\appendix
\section{Bounds on the diamond norm}\label{APPB} 
The lower bound in Eq.~(\ref{diamondnatural}) is a direct consequence of the
definition of the diamond norm~\cite{DIAMOND,DIAMOND1,WatrousBOOK}.
To prove the upper bound of (\ref{diamondnatural}) let us observe that introducing the Schmidt decomposition of the state $|\psi\rangle$ of $Q$ and $Q'$, 
 $|\psi\rangle:= \sum_{j=1}^{M_Q}  \lambda_j  |j\rangle \otimes |j\rangle$,  we can write 
   \begin{eqnarray} 
&& 2 D\big((\Phi\otimes {\cal I}) (|\psi\rangle\langle \psi|),
 (\Phi_{DP}^{(0)} \otimes {\cal I})(|\psi\rangle\langle \psi| ) \big) \nonumber \\
 &&=     \| \sum_{j,j'=1}^{M_A} \lambda_j \lambda_{j'} (\Phi-\Phi_{DP}^{(0)})[ |j\rangle\langle j'|] \otimes |j\rangle\langle j'| \|_1 \nonumber \\
 &&\leq     \sum_{j,j'=1}^{M_A} \lambda_j \lambda_{j'}   \| (\Phi-\Phi_{DP}^{(0)})[ |j\rangle\langle j'|] \otimes |j\rangle\langle j'| \|_1\nonumber \\
 &&\leq    \sum_{j,j'=1}^{M_A} \lambda_j \lambda_{j'}   \| (\Phi-\Phi_{DP}^{(0)})[ |j\rangle\langle j'|]\|_1   \nonumber \\ 
  &&\leq   2    \| \Phi-\Phi_{DP}^{(0)}\|_1   \big(\sum_{j=1}^{M_A} \lambda_j \big)^2 \leq 2  M_A  \| \Phi-\Phi_{DP}^{(0)}\|_1  \;, \nonumber \\ \label{QUESTA1} 
 \end{eqnarray} 
where first we used the convexity of the trace-distance, then the fact that for all $|j\rangle$,  $|j'\rangle$
we have 
\begin{eqnarray} 
\| (\Phi-\Phi_{DP}^{(0)})[ |j\rangle\langle j'|]\|_1 \leq 2 \| \Phi-\Phi_{DP}^{(0)}\|_1 \;,
\end{eqnarray} 
and finally the Chauchy-Schwarz inequality and the normalization condition for the Schmidt coefficients. 
Replacing hence~(\ref{QUESTA1}) into (\ref{defdiam}), Eq.~(\ref{diamondnatural}) finally follows.

\end{document}